\documentclass[preprintnumbers]{revtex4}
\UseRawInputEncoding
\usepackage{amssymb}
\usepackage{amsmath}
\usepackage{graphicx}
\usepackage{dcolumn}
\usepackage{bm}
\usepackage{subfigure}
\usepackage{color}

\begin{document}

\title{Phase structure of the de Sitter Spacetime with KR field based on the Lyapunov exponent}
\author{Yun-Zhi Du,$^{1,2}$ Huai-Fan Li,$^{1,2}$ Yu-Bo Ma,$^{1,2}$ Qiang Gu$^{2}$}
\address{$^1$Department of Physics, Shanxi Datong University, Datong 037009, China\\
$^2$Institute of Theoretical Physics, Shanxi Datong University, Datong, 037009, China }

\thanks{\emph{e-mail:duyzh22@sxdtdx.edu.cn, huaifan999@163.com, yuboma.phy@gmail.com, gudujianghu23@163.com}}

\begin{abstract}
Since the spontaneously broken of the Lorentz symmetry in the gravity theory with the non-minimally coupling between the Kalb-Ramond (KR) field (that acquires a nonzero vacuum expectation value) and the Einstein gravity, there exists the exactly static and spherically symmetric black holes solutions related with the Lorentz violating parameter. Based on this, we consider the corresponding black hole solution in the de-Sitter (dS) spacetime with the KR field and investigate the thermodynamic properties in the expanded phase space through introducing the interplay entropy between the black hole and cosmological horizons. Especially we analyze the effect of the Lorentz-violating parameter on the thermodynamic properties. Furthermore, the Lyapunov exponent and the shadow of these static and spherically symmetric black holes in this Lorentz-violating gravity theory are also investigated. These study will open a new perspective to probe the thermodynamics of black holes.
\end{abstract}

\maketitle

\section{Introduction}

The black hole thermodynamics which can reveal the nature and properties of black holes has been paid more and more attention to probe the profound relationship between general gravity, quantum field theory, thermodynamics, and statistical mechanics. It can provide guidelines for harmonizing the quantum theory and gravitational theory. In the seventies, Bekenstein and Hawking had discovered that the black hole in the AdS spacetime is not only a gravitational system, but also is a thermodynamic system and obeys the four laws of the black hole thermodynamics \cite{Bekenstein1974,Bardeen1973}. That makes lots of investigations about the thermodynamics of black holes in AdS spacetime \cite{Hawking1983,Cai2008,Wei2015,Zhang2015,Du2021,He2010,Yao2022,Wei2022,He2017,Huang2021,Bai2022,Wang2020,Hamil2023} arise. Motivated by the recent investigation of the striking similarity between the thermodynamic phase structure of AdS black holes in the canonical ensemble and that of the van der Waals system, the physics of the charged AdS black holes were explored in more detail \cite{Chamblin1999, Chamblin1999a}. Subsequently, Mann discovered that the first law of thermodynamics at the black hole horizon and the cosmological horizon still holds on in a dS spacetime \cite{Dolan2013}, which induces some investigations on black holes in dS spacetime \cite{Tannukij2020,Chabab2021,Wei2019,Simovic2019,Nam2018,Kubiznak2016,Zhang2014,Cvetic2002}. There are some works that had been done to treat the black hole and its surrounding radiation field as a thermodynamic system, and it has been found to have thermodynamic properties similar to those of AdS black holes \cite{Zhao2021,He2018,Ma2018,Zhao2018,Zhang2019,Liu2019,Guo2020}.
However compared to ordinary thermodynamic systems, black holes are quantum in nature, and they are reflecting a statistically significant macroscopic quantum effect. Therefore, investigating the thermodynamic properties of AdS black holes and the dS spacetime with black holes will provide a thermodynamic basis for our understanding of quantum gravity. Additionally inspired by the duality of AdS/CFT \cite{Gubser1998,Maldacena1998} and dS/CFT \cite{Strominger2001}, the properties of strongly coupled conformal field theory can be traced by the thermodynamics of AdS black holes and dS spacetimes with black hole. For example, Hawking and Page \cite{Hawking1983} found the famous thermodynamic phase transition known as the Hawking-Page (HP) phase transition, which exists between the thermal radiation phase and a stable large black hole phase in AdS spacetime as the temperature increases. That makes the phase structure of AdS black holes and dS spacetimes with different black hole become more charming. On the other hand, since our universe at the time of the explosion was a quasi-dS spacetime, this has prompted further research into the dS spacetime with black holes and their thermodynamic properties.

As so far, although Einstein's general relativity which is the extension the principles of special relativity and maintains the local Lorentz symmetry, has been experimentally verified, there are some theories (such as the string theory \cite{Kostelecky1989}, loop quantum gravity \cite{Alfaro2002}, Horava-Lifshitz gravity \cite{Horava2009},
and so on) that propose the possibility of the Lorentz symmetry breaking (LSB) at certain energy scale and provide valuable insights into the nature of spacetime. In order to studying the LSB many extensions of the Standard Model had emerged, the simplest of which is the bumblebee model \cite{Kostelecky1989,Kostelecky1989a,Kostelecky1989b,Bailey2006,Bluhm2008}.
Firstly, the bumblebee field was a vector field which acquires a nonzero vacuum expectation value (VEV) and leads to the violation of particle local Lorentz invariance. In this frame there were a series investigations of different black holes \cite{Bluhm2008,Casana2018,Kanzi2019,Maluf2021,Cai2023,Xu2023,Ding2020,Ding2021,Wang2022,Liu2023}.
On the other hand, instead of a vector field, a rank-two antisymmetric tensor field called the Kalb-Ramond (KR) field was introduced as the bumblebee field to act the source of LSB \cite{Altschul2010}. As the KR field nonminimally couples to gravity and aquires a nonzero VEV, the Lorentz symmetry is spontaneously broken. A static and spherically symmetric solution was proposed under the VEV background of KR field \cite{Lessa2020} and the corresponding investigations of this solution were given in \cite{Atamurotov2022,Kumar2020}. Recently, the authors \cite{Yang2023} have propose another new exact solution for the static and spherically symmetric spacetime with the cosmological constant under the nonzero VEV background of KR field. In this work, we will consider the dS spacetime with the KR field under the nonzero VEV background.

As we know that in order to study the nonlinear systems, the chaos theory was established to discover ordered structures and laws hidden in some seemingly disordered phenomenas (such as the fractals and the sensitivity of the initial value of a dynamical system known as the butterfly effect). The authors in \cite{Suzuki1997,Lu2018,Hartl2003,Bombelli1992} had studied the perturbations and particle orbits of black hole systems through the chaos theory. The Lyapunov exponents were introduced as an indicator of the separation rate between neighboring trajectories to explore the sensibility of these systems to the initial condition. The results showed that the positive Lyapunow exponent means that the system is of chaotic property, i.e., an initially slight difference will lead to the exponential separation of trajectories. As Lyapunov exponent is zero, the system is stable, and the neighboring trajectories keeps at a certain distance. When the Lyapunov exponent is negative, the particle orbits will be asymptotic stability and the trajectories nearby will tend to overlap. Therefor the Lyapunov exponent was applied to probe the orbits stability and rate of orbits divergence of the massive and massless particles in the outer space of black holes \cite{Wang2017,Chen2016,Yang2023a}. For the spherically symmetric black hole (for example, Schwarzschild black hole), its shadow as a black circular disk can be shown in \cite{Synge1966}. Subsequently, it is accompanied with many other studies on black hole shadows \cite{Vries1999,Bardeen1972,Grenzebach2014,Guo2018,Hennigar2018,Amir2018,Jusufi2020}. The black hole shadows can reveal many characteristics of black holes. Thus, accounting to the observational effect of the Lyapunov exponents, we are looking forward to explore the relationship between phase transition, shadow of black hole and the Lyapunov exponents. And we hope to observe the chaotic properties of the orbits and  the thermodynamic of black holes, when the thermodynamic properties are investigated with its Lyapunov exponent.

The work is organized as follows. Firstly, we would like to discuss the conditions for the existence of black hole and cosmological horizons in the KR-dS spacetime, as well as the influence of the Lorentz violating parameter, and give the range of the position ratio $x$ of two horizons. Then the effective thermodynamic quantities and the critical point of the KR-dS spacetime are presented. Furthermore we also analyze the behaviors of the heat capacities and the first-order phase transition. Finally, based on the effective thermodynamic quantities and the method of Hamilton, we discuss the Lyapunov exponent of these static and spherically symmetric black holes under the KR-dS spacetime background, so that to probe the phase structure of this system. And the light rays and the shadow of KR-dS black hole are investigated to explore the effect of Lorentz violation on the motion of photon near the
black hole. The Sec. \ref{Seven} is the summary.

\section{The exact solutions of the KR black hole}
\label{two}
For the four-dimensional Lorentz-violating gravity theory with the non-minimally coupling between the Kalb-Ramond (KR) field and the Einstein-Hilbert action, the action reads \cite{Altschul2010,Lessa2020,Yang2023,Duan2023}
\begin{eqnarray}
S=\frac{1}{16\pi G}\int d^4x\sqrt{-g}\left[R-2\Lambda-\frac{1}{6}H^{\mu\nu\rho}H_{\mu\nu\rho}-V\left(B^{\mu\nu}B_{\mu\nu}\right)+\xi_2B^{\rho\mu}B^{\nu}_\mu R_{\rho\nu}+\xi_3B^{\mu\nu}B_{\mu\nu}R\right]+\int d^4x\sqrt{-g}L_M,
\end{eqnarray}
where $G$ is the Newtonian constant of gravitation, $\xi_{2,3}$ are the coupling constants, $H_{\mu\nu\rho}\equiv\partial_{[\mu}B_{\nu\rho]}$ is the strength of KR field, and $\Lambda$ is the cosmological constant. The existence of the potential function $V\left(B^{\mu\nu}B_{\mu\nu}\right)$ that relates to $B^{\mu\nu}B_{\mu\nu}$ is to maintain the invariant of this theory with the local Lorentz transformations.

For a static and spherically symmetric spacetime in which the non-minimally coupled KR field is of the nonzero vacuum expectation value (VEV), there exists three types of RN-like black hole solutions in the asymptotically flat, AdS, and dS spacetimes, the corresponding solution read
\begin{eqnarray}
f(r)=\left\{
  \begin{array}{ll}
\frac{1}{1-a}-\frac{2M}{r}+\frac{Q^2}{(1-a)^2r^2} & ~~~~~~~~~~~\text{$\Lambda=0$} \\
    \frac{1}{1-a}-\frac{2M}{r}+\frac{Q^2}{(1-a)^2r^2}-\frac{\Lambda r^2}{3(1-a)} & ~~~~~~~~~~~\text{$\Lambda\neq0$}~~~~~
  \end{array}\right.,
\end{eqnarray}
where the dimensionless parameter $a$ is the Lorentz-violating parameter and characterizes the effect of the Lorentz violation caused by the nonzero VEV of the KR field. Since the strong constraints from some classical gravitational experiments within the Solar System, the Lorentz-violating parameter is supposed to be very small \cite{Yang2023}. Furthermore, the Lorentz-violating efffecs cannot be eliminated by the coordinate transformations, while the horizon singularity can be removed through a coordinate transformation, the center singularity is an intrinsic and nonremovable singularity \cite{Yang2023}. When the cosmological constant is negative, $\Lambda<0$, the black hole event horizon exists for any parameter, and thus the above metric always supports black hole solutions in this case. However, when the cosmological constant is positive, $\Lambda>0$, the black hole solutions exist only for the parameters $a$, $Q$, $\Lambda$, and $M$ satisfying $(1-a)\Lambda \left[8Q^4-36(1-a)^3Q^2M^2+27(1-a)^6M^4\right]+4Q^6\Lambda^2\leq(1-a)^2\left[3(1-a)^3M^2-Q^2\right]$, where
the equality sign represents the case when the event horizon and the cosmological horizon coincide with each other. As $a=0$, the solution reduces to the standard RN one. In the following, we will mainly focus on the KR black hole solution in a dS spacetime, so we call the system as the KR-dS spacetime with a RN-like black hole.

For this Lorentz-violating gravity system in the KR-dS spacetime, there are three horizons: the inner and outer horizon of the black hole, the cosmological horizon, which satisfy the expression $f(r_-,r_+,r_c)=0$. The mass parameter can be obtained from the above equation
\begin{eqnarray}
M=\frac{r_{+,c}}{2}\left(\frac{1}{1-a}+\frac{Q^2}{(1-a)^2r_{+,c}^2}-\frac{\Lambda r_{+,c}^2}{3(1-a)}\right), \label{M}
\end{eqnarray}
where $M$ is the mass parameter of black hole. For the KR-dS spacetime with the RN-like black hole, there exist the local minimum ($M_{min}$) and local maximum ($M_{max}$) for black hole mass. As $M_{min}\leq M_0\leq M_{max}$, this system have three horizons, i.e., the inner and outer horizons of black hole, the cosmological horizon. When the local minimum and local maximum merge into an inflexion point, these three horizons coincide together. In this case, the hole black is called the ultracold black hole ($r_{ucold}=\sqrt{\frac{2}{1-a}}Q,~\Lambda Q^{2}=\frac{1-a}{4},~M_{ucold}=\frac{2Q}{3(1-a)}\sqrt{\frac{2}{1-a}}$). Generally, the black hole horizons do not coincide with the cosmological one, thus the conditions $4\Lambda Q^{2}\leq 1-a $ and $M_{min}\geq M_{ucold}$ should be held on. From eq. (\ref{M}) and the definition $x\equiv r_+/r_c$, we have
\begin{eqnarray}
\Lambda=\frac{x^2}{1+x+x^2}\frac{3}{r^2_+}-\frac{x^3}{1+x+x^2}\frac{3Q^2}{(1-a)r^4_+}.\label{ql2}
\end{eqnarray}
When $M_0=M_{min}>M_{ucold}$: the black hole inner and outer horizons coincide together, such black hole is called the cold one. As $M_0=M_{max}>M_{ucold}$: the black hole and cosmologic horizons coincide together ($x=1$), this black hole is called the Nariai one, by solving the above equation we have
\begin{eqnarray}
\frac{Q^2}{r_+^2}=\frac{1-a}{2}\left(1+\sqrt{1-\frac{4\Lambda Q^2}{1-a}}\right).\label{phimax}
\end{eqnarray}
Generally, the ratio between two horizons should be from a minimum instead of zero to one. For the Nariai black hole, substituting eq. (\ref{phimax}) into eq. (\ref{ql2}), the minimum of $x$ is constrained by the following expression
\begin{eqnarray}
\frac{x_{min}^2}{1+2x_{min}+3x_{min}^2}=\frac{1}{6}\left(1-\sqrt{1-\frac{4\Lambda Q^2}{1-a}}\right).\label{x}
\end{eqnarray}
In this system, the KR-dS black hole can be survives with the range of $x_{min}\leq x\leq1$.

\section{Effective thermodynamic quantities and phase transition}
\label{three}

For a dS spactime with a black hole, one can describe the thermodynamical property on the black hole and cosmological horizons, respectively, where the thermodynamical laws are still satisfied. The thermodynamic quantities on two horizons of the KR-dS spacetime with a RN-like black hole are
\begin{eqnarray}
T_c&=&-\frac{(1-x)^2}{4\pi(1-a)(1-x^3)r_c}\left(1+2x-\frac{Q^2(3+2x+x^2)}{(1-a)xr_c^2}\right),~~
S_c=\pi r_c^2,~~~V_c=\frac{4\pi r_c^3}{3},\\
T_+&=&\frac{(1-x)^2}{4\pi(1-a)(1-x^3)r_+}\left(1+2x-\frac{Q^2(3+2x+x^2)}{(1-a)xr_+^2}\right),~~
~S_+=\pi r_+^2,~~~V_+=\frac{4\pi r_+^3}{3}.
\end{eqnarray}
When the hawking radiation temperatures on two horizons are equal to each other, i.e., in the lukewarm case the potentials on two horizons become
\begin{eqnarray}
\frac{Q^2}{r_+^2}=\frac{1-a}{(1+x)^2},
\end{eqnarray}
and the same radiation temperature on two horizons reads
\begin{eqnarray}
T=T_c=T_+=\frac{x(1-x)}{2\pi r_+(1-a)(1+x)^2}.
\end{eqnarray}
For the KR-dS spacetime with the RN-like black hole, we mainly investigate the space between black hole and cosmological horizons, i.e., the thermodynamic volume of this system reads
\begin{eqnarray}
V=V_c-V_+=\frac{4\pi(1-x^3)r_+^3}{3x^3}.\label{V}
\end{eqnarray}
The boundary of our considered system are two horizons which have different radiation temperatures. That is because the system is in gravity field, and the horizon temperatures are deformed by the warped space. Thus we regard this system as an ordinary thermodynamic system in the thermodynamic equilibrium, which is of the thermodynamic quantities ($T_{eff},~P_{eff},~V,~S$). Here we point out the entropy is not only the sum of two horizons, it should contain the connection between two horizons which arise from the gravity effect. We assume the total entropy of this system as the following form
\begin{eqnarray}
S=\frac{\pi(1+x^2+f_0(x))r_+^2}{x^2},\label{S}
\end{eqnarray}
where $f_0(x)$ stands for the interplay between two horizons and it arises from the gravity effect. In the following, we mainly focus on the process of obtaining the function $f_0(x)$. Since the thermodynamic quantities of this system satisfy the first law
\begin{equation}
d M=T_{e f f} d S-P_{e f f} d V+\Phi_{e f f} d Q,
\end{equation}
thus the effective temperature and pressure can be obtained from the following expressions
\begin{eqnarray}
T_{e f f}&=&\frac{\partial{M}}{\partial{S}}\bigg|_{{V},Q}
=\frac{\frac{\partial{M}}{\partial r_+}\frac{\partial{V}}{\partial x}-\frac{\partial{M}}{\partial x}\frac{\partial{V}}{\partial r_+}}{\frac{\partial{S}}{\partial r_+}\frac{\partial{V}}{\partial x}-\frac{\partial{S}}{\partial x}\frac{\partial{V}}{\partial r_+}}\bigg|_{Q},\label{Teff}\\
P_{e f f}&=&-\frac{\partial{M}}{\partial{V}}\bigg|_{{S},Q}
=\frac{\frac{\partial{M}}{\partial r_+}\frac{\partial{S}}{\partial x}-\frac{\partial{M}}{\partial x}\frac{\partial{S}}{\partial r_+}}{\frac{\partial{V}}{\partial x}\frac{\partial{S}}{\partial r_+}-\frac{\partial{V}}{\partial r_+}\frac{\partial{S}}{\partial x}}\bigg|_{Q}.\label{Peff}
\end{eqnarray}
As considering eq. (\ref{ql2}), the temperature corresponding to two horizons is equal to each other. In this case, the effective temperature should have the same form
\begin{eqnarray}
T_{e f f}=\frac{x(1+x^4){T}_+}{(1-x^3)\left[x(1+x)+x^2f_0+\frac{1}{2}(1-x^3)f_0'\right]}={T}_+.
\end{eqnarray}
As $x=0$, the interplay between two horizons should be vanished, i.e., $f_0(0)=0$. By solving above equation the interaction function of entropy reads
\begin{eqnarray}
f_0(x)=\frac{8}{5}(1-x^3)^{\frac{2}{3}}-\frac{2(4-5x^3-x^5)}{5(1-x^3)}.\label{f0}
\end{eqnarray}
Substituting eq. (\ref{f0}) into eqs. (\ref{Teff}) and (\ref{Peff}), the effective temperature and effective pressure are satisfied the following forms
\begin{eqnarray}
T_{eff}&=&\frac{1}{(1-a)r_+f_1(x)}\left[f_2(x)-\frac{Q^2 f_3(x)}{(1-a)r_+^2}\right],\label{T}\\
P_{eff}&=&-\frac{1}{(1-a)r_+^2f_4(x)}\left[f_5(x)+\frac{Q^2 f_6(x)}{(1-a)r_+^2}\right]\label{P}
\end{eqnarray}
with
\begin{eqnarray}
f_1(x)&=&\frac{4\pi(1+x^4)}{1-x},~~f_2(x)=1+x-2x^2+x^3+x^4,f_3(x)=(1+x+x^2)(1+x^4)-2x^3,~~\nonumber\\
f_4(x)&=&\frac{8\pi(1+x^4)}{x(1-x)},~~~
f_5(x)=x(1+x)(x+f_0'/2)-\frac{(1+2x)(1+x^2+f_0)}{1+x+x^2},~~~\nonumber\\
f_6(x)&=&\frac{(1+2x+3x^2)(1+x^2+f_0)}{1+x+x^2}-x(1+x)(1+x^2)(x+f_0'/2).\nonumber
\end{eqnarray}

With the following critical condition of the effective pressure
\begin{eqnarray}
\frac{\partial P_{eff}}{\partial V}\bigg|_{T_{eff}}=\frac{\partial^2P_{eff}}{\partial V^2}\bigg|_{T_{eff}}=0,
\end{eqnarray}
and eqs. (\ref{V}), (\ref{f0}), (\ref{T}), and (\ref{P}), we can get influence of parameters $Q$ and $a$ on the phase transition in Tabs. \ref{Tab1} and \ref{Tab2}. It is especially interesting that the critical ratio, horizon radius, volume, and entropy are only related to the charge parameter, instead of the Lorenz-violating parameter, while the critical effective temperature and pressure are determined by both two parameters $Q$ and $a$. Furthermore, for the system with a fixed charge parameter both the critical effective temperature and pressure are increasing with the Lorentz-violating parameter $a$, while they are decreasing with the charge parameter $Q$ as the system with a fixed Lorentz-violating parameter.

\begin{table}[htbp]
\centering
\caption{The thermodynamic quantities at the critical point with $Q=1$: $x_c=0.656434,~r_+^c=2.64432,~V^c=196.364,~S^c=68.2395.$}
\begin{tabular}{c|c|c|c|c|c|c}
\hline\hline
\centering
Lorentz-violating parameter~~&$a=0$~~& $a=0.0001$~~&$a=0.001$~~&$a=0.01$~~&$a=0.1$~~&$a=0.2$ \\ \hline
$T_{eff}^c$~~&$0.00863395$~~& $0.00863481$~~&$0.00864259$~~&$0.00872116$~~&$0.00959328$~~&$0.0107924$ \\ \hline
~~$P_{eff}^c$ ~~&0.00058368~~&0.000583745~~&0.000584271~~&0.000589582~~&0.00064854~~&0.000729608\\ \hline
$\Lambda$~~&0.0802544~~&0.0802536~~&0.0802461~~&0.0801704~~&0.0793306~~&0.0781757\\ \hline\hline
\end{tabular}\label{Tab1}
\end{table}

\begin{table}[htbp]
\centering
\caption{The thermodynamic quantities at the critical point with $Q=0.85$: $x_c=0.666554,~r_+^c=2.48271,~V^c=152.351,~S^c=58.987.$}
\begin{tabular}{c|c|c|c|c|c|c}
\hline\hline
\centering
Lorentz-violating parameter~~&$a=0$~~& $a=0.0001$~~&$a=0.001$~~&$a=0.01$~~&$a=0.1$~~&$a=0.2$ \\ \hline
$T_{eff}^c$~~&$0.0093246$~~& $0.00932554$~~&$0.00933394$~~&$0.00941879$~~&$0.0103607$~~&$0.0116558$ \\ \hline
~~$P_{eff}^c$ ~~&0.000625542~~&0.000625605~~&0.000626168~~&0.000631861~~&0.000695047~~&0.000781928\\ \hline
$\Lambda$~~&0.0944389~~&0.0944381~~&0.0944308~~&0.094358~~&0.0935495~~&0.0924379\\ \hline\hline
\end{tabular}\label{Tab2}
\end{table}

\begin{figure}[htp]
\subfigure[$~~C_{P_{eff}}-T_{eff}$]{\includegraphics[width=0.4\textwidth]{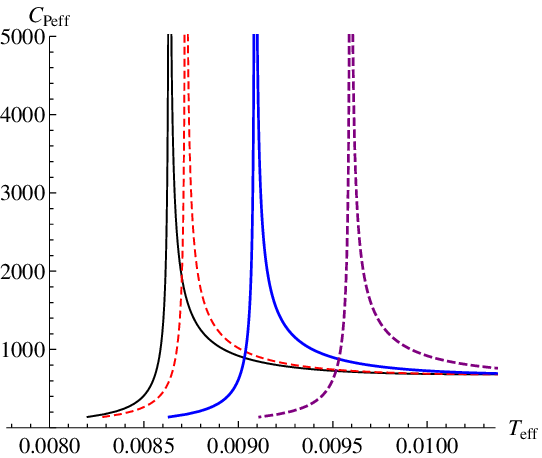}\label{Cp}}~~~~
\subfigure[$~~C_V-T_{eff}$]{\includegraphics[width=0.4\textwidth]{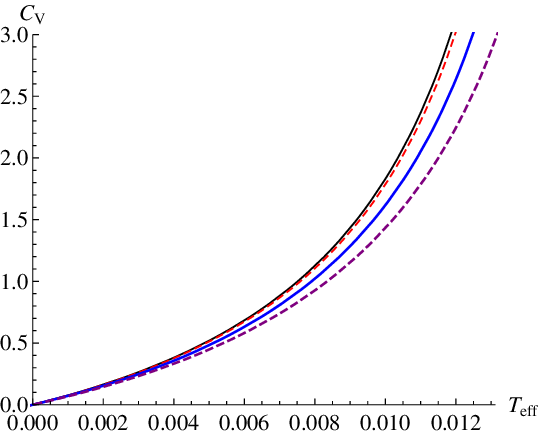}\label{Cv}}
\caption{The behaviors of $C_{P_{eff}}$ and $C_V$. The parameters are set to $Q=1$, $a=0$ (black thin lines), $a=0.01$ (red thin dashed lines), $a=0.05$ (blue thick lines), and $a=0.1$ (purple thick dashed lines).}\label{CPV}
\end{figure}

When the system undergoing a isobaric, isothermal, and isovolume processes, respectively, from eqs. (\ref{T}), (\ref{P}), and (\ref{V}) we can obtain the physical black hole horizon radius as
\begin{eqnarray}
r_p&=&\sqrt{\frac{-f_5+\sqrt{f_5^2-4Q^2P_{eff}f_4f_6}}{2(1-a)P_{eff}f_4}},~~~~~~
r_v=\left(\frac{4\pi[1-x^3]}{3x^3V}\right)^{1/3},\label{rpv}\\
r_t&=&\frac{\frac{Q}{2(1-a)}\sqrt{\frac{3f_3}{f_1}}}{\cos\left(\theta+\frac{4\pi}{3}\right)},~~~~~~
\theta=\frac{1}{3}\arccos\left(\frac{-QT_{eff}f_1\sqrt{27(1-a)f_2f_3}}{2f_2^2}\right).\label{rt}
\end{eqnarray}
In order to analyze the effect of the Lorentz-violating parameter on the stability of this system, we investigate the behaviours of the heat capacities at the constant volume and the effective pressure, which read
\begin{eqnarray}
C_{P_{eff}}=T_{eff}\frac{\frac{\partial S}{\partial r_+}\frac{\partial P_{eff}}{\partial x}-\frac{\partial S}{\partial x}\frac{\partial P_{eff}}{\partial r_+}}{\frac{\partial T_{eff}}{\partial r_+}\frac{\partial P_{eff}}{\partial x}-\frac{\partial T_{eff}}{\partial x}\frac{\partial P_{eff}}{\partial r_+}},~~~~
C_V=T_{eff}\frac{\frac{\partial S}{\partial r_+}\frac{\partial V}{\partial x}-\frac{\partial S}{\partial x}\frac{\partial V}{\partial r_+}}{\frac{\partial T_{eff}}{\partial r_+}\frac{\partial V}{\partial x}-\frac{\partial T_{eff}}{\partial x}\frac{\partial V}{\partial r_+}}.
\end{eqnarray}
With the above equations and eqs. (\ref{rpv}), (\ref{rt}), the corresponding curves of the heat capacities are demonstrated in Fig. \ref{CPV}, which indicate that the system is stable. It is very unique that the heat capacity at the constant volume is no-zero, which is fully different from AdS black hole systems. This can be regarded as a significant difference between AdS black holes and dS spacetimes with black hole.

\begin{figure}[htp]
\subfigure[$~~P_{eff}-V$]{\includegraphics[width=0.4\textwidth]{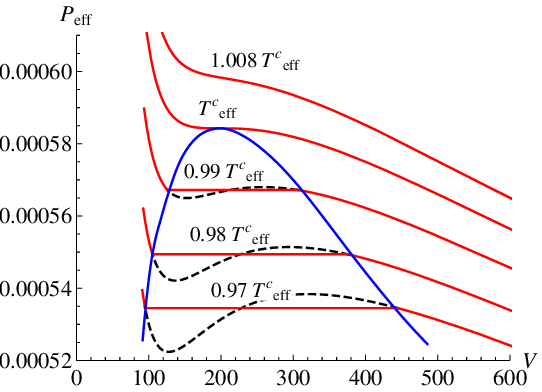}\label{PV-KR}}~~~~
\subfigure[$~~G_{T_{eff}}-P_{eff}$]{\includegraphics[width=0.4\textwidth]{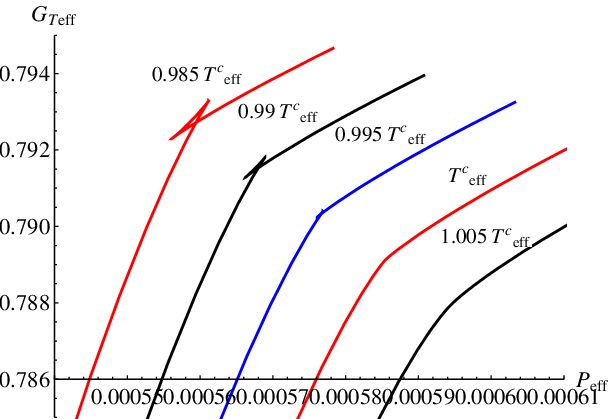}\label{GP-KR}}~~~~
\caption{The phase diagrams of $P_{eff}-V$ and $G_{T_{eff}}-P_{eff}$ nearby the critical point with $a=0.0001$ and $Q=1$.}\label{PVGP}
\end{figure}

\begin{figure}[htp]
\includegraphics[width=0.4\textwidth]{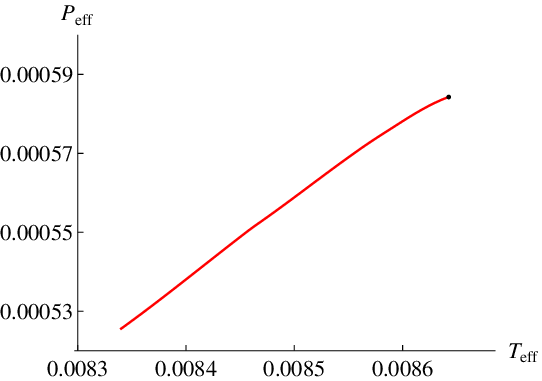}
\caption{The coexistent curve of $P_{eff}-T_{eff}$ nearby the critical point with $a=0.0001$ and $Q=1$.}\label{P0T0-KR}
\end{figure}

On the other hand, the Gibbs free energy of this system can be expressed as
\begin{eqnarray}
G=M-T_{eff}S+P_{eff}V.
\end{eqnarray}
To demonstrate the properties of the KR-dS spacetime with a RN-like black hole, we present the diagrams of $P_{eff}-V$ and the Gibbs free energy against the effective pressure in Fig. \ref{PVGP}. As shown in Fig. \ref{PV-KR} when $T_{eff}\leq T_{eff}^c$, there are two extreme points and one inflection point, whose thermodynamic properties of the KR-dS spacetime with the RN-like black hole are similar to the Reissner-Nordstrom-AdS black holes. There's a phase transition temperature between extremely large and extremely small values according to the Maxwell's law of equal area. As $T_{eff}= T_{eff}^c$, it is in a state of criticality with the two extreme points converging. For $T_{eff}> T_{eff}^c$, no extreme point exists, which means no phase transition will occur. From Fig. \ref{GP-KR}, we can see that for $T_{eff}\leq T_{eff}^c$, there is a characteristic swallow tail composed of three phases i.e., small volume phase, intermediate volume phase, and large volume phase. As $T_{eff}= T_{eff}^c$ the system is accompanied by a second order phase transition; and when $T_{eff}>T_{eff}^c$ the Gibbs free energy decrease monotonically with the effective pressure, which shows no phase transition exists. Furthermore, we also present the coexistent curve of the first-order phase transition for this system in Fig. \ref{P0T0-KR}. As shown in \cite{Yang2023}, through several classical gravitational experiments within the Solar System, including
the perihelion precession of Mercury, deflection of light, and Shapiro time delay the Lorentz-violating effect did contribute to the corrections observed in these experiments, which had given a certain constrain the value of the Lorentz-violating parameter: $-10^{-13}<a<10^{-10}$. Since the small value of the Lorentz-violating parameter, here we only present one curve of the coexistent phases in the first-order phase transition with $a=0.0001$.

\section{Lyapunov exponent and optical properties}
\label{four}
The Lyapunov exponent, which is an important method to research the sensibility and complexity of dynamic systems, can reflect whether the orbital near a black hole is divergence and convergence in the phase space. As shown in Refs. \cite{Guo2022,Cardoso2009,Wei2023,Yang2023a}, a positive Lyapunov exponent stands for a divergence of nearby geodesics around black holes. Lyapunov exponent has also been applied to investigate the large/small black hole phase transition of the RN-AdS black hole \cite{Cardoso2009,Wei2023}, the Born-Infeld AdS black hole \cite{Yang2023a}, and the $D$-dimensional charged Gauss-Bonnet AdS black holes \cite{Lyu2023}. In this part we will probe the stability of this system by the Lyapunow exponent, and investigate the influence of the Lorentz-violating parameter on the optical properties of the KR-dS black hole.

\begin{figure}[htp]
\includegraphics[width=0.4\textwidth]{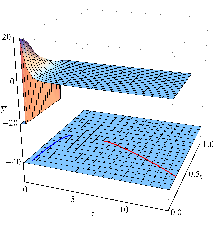}
\caption{The effective potential respect $\bar V$ to the radial radius and the ratio between two horizons. The parameters are set to $a=0.0001$, $Q=2$, $b=1$, and $T_{eff}=0.85T_{eff}^c$.}\label{Vrrx-KR}
\end{figure}

The massless photon along its geodesics is governed by the Lagrangian density
\begin{eqnarray}
H=-\frac{1}{2}g_{\mu\nu}g^{\mu\nu}=0.
\end{eqnarray}
Since the spherically symmetric of this system, we will pay attention on the equatorial geodesics with $\theta=\pi/2$. Thus the Lagrangian becomes
\begin{eqnarray}
-f(r)\left(\frac{dt}{d\tau}\right)^2+f^{-1}(r)\left(\frac{dr}{d\tau}\right)^2
+r^2\left(\frac{d\phi}{d\tau}\right)^2=0,\label{H0}
\end{eqnarray}
where $\tau$ is the proper time. Furthermore, there are two conserved quantities of the static and spherically symmetric spacetime
\begin{eqnarray}
E=-\frac{\partial H}{\partial \dot t}=f(r)\dot t,~~~~L=\frac{\partial H}{\partial \dot\phi}=r^2\dot\phi. \label{EL}
\end{eqnarray}
With the above equations, we have
\begin{eqnarray}
\dot r^2=E^2-\frac{L^2f(r)}{r^2}\equiv V_{eff}. \label{Vr}
\end{eqnarray}
Substituting eqs. (\ref{M}), (\ref{ql2}), and (\ref{rt}) into the above equation, we have the effective potential $\bar V\equiv\frac{V_{eff}}{E^2}$ as the following form
\begin{eqnarray}
&&\bar V=1-\frac{b^2}{r^2}\nonumber\\
&&~~\left[\frac{1}{1-a}\left(1-\frac{(1-x^2)r_+}{(1-x^3)r}\right)-\frac{Q^2}{(1-a)^2r^2}
\left(1-\frac{(1-x^4)r}{(1-x^3)r_+}\right)+\frac{xr^2}{(1-a)(1+x+x^2)r_+^2}\left(1-\frac{x}{(1-a)r_+^2}\right)\right]
\end{eqnarray}
with the impact parameter $b\equiv \frac{L}{E}$, where $r$ is the radius of the photon orbit. The effective potential is a complicated function affected by the parameters horizon $r_+$, the ratio between two horizons $x$, the impact parameter $b$, the Lorentz-violating parameter $a$, and the charge parameter $Q$. Here we exhibit its property by numerically obtaining a diagram of the effective potential as the function of the $r$ and $x$ for the given parameters $Q,~a,~b,~T_{eff}$ in Fig. \ref{Vrrx-KR}. The blue curve that projected onto $\bar V=-40$ is the extreme points of the effective with $\bar V'=0$ and $\bar V''<0$, while the red curve on the same plane represent the ones when $\bar V'=0$ and $\bar V''>0$. They are corresponding to unstable equilibria and the stable equilibria respectively.

The Lyapunov exponent of the null geodesic reads \cite{Cardoso2009,Wei2023,Yang2023a}
\begin{eqnarray}
\lambda=\pm\sqrt{\frac{V''_{eff}}{2\dot t^2}}, \label{lambda}
\end{eqnarray}
where the prime stands for the derivative of the radial radius $r$. One can probe the stability of the photon circular orbital by the second derivative on the effective potential under the conditions $V_{eff}=0$ and $V'_{eff}=0$. Because the signs of Lyapunov exponent has nothing to do with the stability of the photon's circular orbital, we will choose the positive value of Lyapunov exponent in the following.

\begin{figure}[htp]
\subfigure[$~Q=1,~a=0.0001,~b^2=0.65,~L^2=50$]{\includegraphics[width=0.32\textwidth]{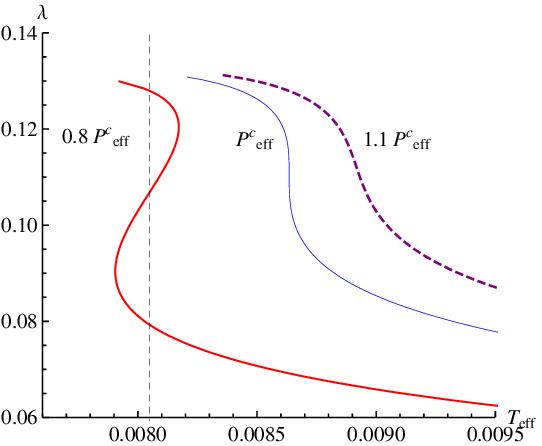}\label{lambda-Teff-P}}~~
\subfigure[$~Q=1,~0.8P_{eff}^c,~b^2=0.65,~L^2=50$]{\includegraphics[width=0.32\textwidth]{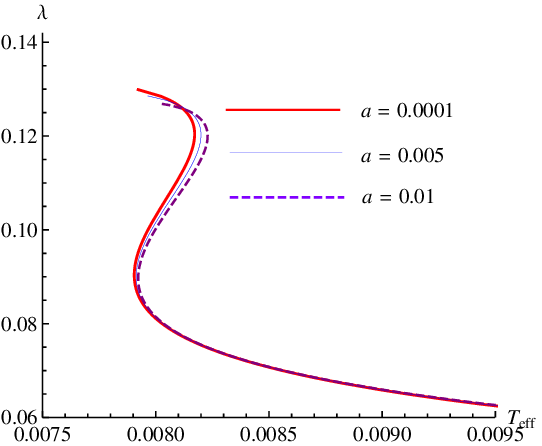}\label{lambda-Teff-a}}~~
\subfigure[$~Q=1,~a=0.0001,~0.8P_{eff}^c,~L^2=50$]{\includegraphics[width=0.32\textwidth]{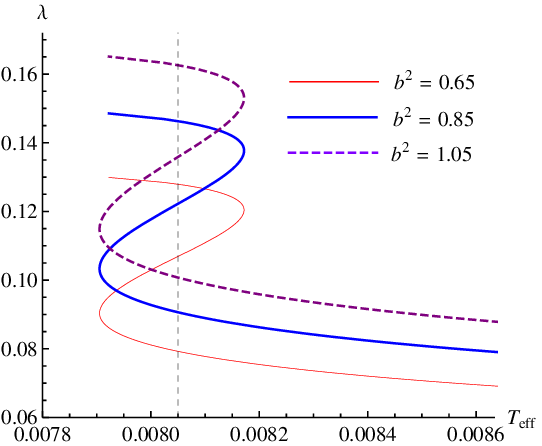}\label{lambda-Teff-b}}\\
\subfigure[$~Q=1,~a=0.0001,~b^2=0.65,~0.8P_{eff}^c$]{\includegraphics[width=0.32\textwidth]{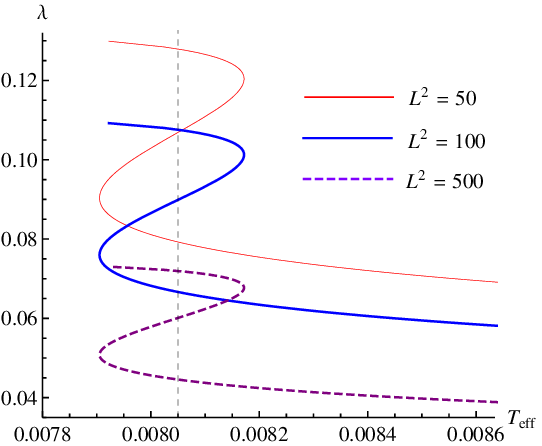}\label{lambda-Teff-L}}~~
\subfigure[$~a=10^{-4},~b^2=0.65,~L^2=50,~0.8P_{eff}^c$]{\includegraphics[width=0.32\textwidth]{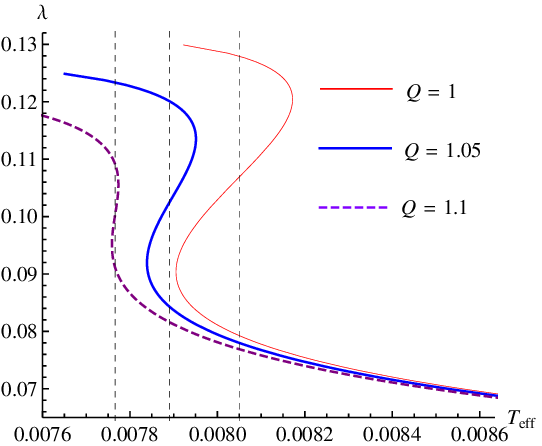}\label{lambda-Teff-Q}}~~
\caption{The Lyapunov exponent of the photon circular orbital vs the different parameters.}\label{lambda-Teff}
\end{figure}

With eqs. (\ref{H0}) and (\ref{EL}), we have the orbit equation for the massless photon,
\begin{eqnarray}
\left(\frac{dr}{d\phi}\right)^2=r^4\left(\frac{E^2}{L^2}-\frac{f(r)}{r^2}\right),\label{drphi}
\end{eqnarray}
which indicates that the orbit equation only dependents on one constant of motion, i.e., the impact parameter $b=L/E$. Note that the above equation is of the same form as an energy conservation law in one-dimensional classical mechanics shown as eq. (\ref{Vr}), where the effective potential also dependents on the impact parameter with the coordinate $\phi$ playing the role of the time variable. With the fixed impact parameter we can visualize the radial motion by a motion in the classical potential. Here the unstable photon radius is marked by $r_{ph}$, which is determined by the implicit express
\begin{eqnarray}
r_{ph}=\frac{3M}{2}\left(1-a+\sqrt{1-\frac{8Q^2}{9M^2(1-a)^3}}\right).
\end{eqnarray}
In order to get a physical photon orbit, the Lorentz-violating parameter should be satisfied the condition: $a\leq1-2/3\left(3Q^2/M^2\right)^{1/3}$. In situations where the light ray approaches the center and then goes out again after reaching a minimum radius, it is convenient to rewrite the orbit equation (\ref{drphi}) using $R$ instead of $b$. As $R$ corresponds to the turning point of the trajectory, the condition $dr /d\phi|_R=0$ should be hold on, and
\begin{eqnarray}
b^2=L^2/E^2=\frac{R^2}{f(R)}.\label{bR}
\end{eqnarray}
Substituting eq. (\ref{EL}) into eq. (\ref{lambda}) and considering the circular orbital conditions, the Lyapunov exponent of the photon circular orbital becomes
\begin{eqnarray}
\lambda=\sqrt{\frac{r_{ph}^2f(r_{ph})V''_{eff}(r_{ph})}{2L^2}}.
\end{eqnarray}
Plugging the effective temperature into the above equation, one can express the Lyapunov exponent in terms of $T_{eff}$. The corresponding behaviors of the Lyapunov exponent against the effective temperature with different parameters of $P_{eff},~a,~b,~L,~Q$ are exhibited in Fig. \ref{lambda-Teff}. For the parameters of $Q,~a,~b,~L$, there are three phases which are coexistent as $P_{eff}<P_{eff}^c$, that indicates the existence of the first-order phase transition. And the Lyapunov exponents of the small- and big-volume phases are both decrease with increasing of the effective temperature, while the Lyapunov exponent for the immediate-volume phase increases with the effective temperature. These results are different from that for AdS black holes \cite{Guo2022}. Furthermore, the point of the first-order phase transition is independent with the parameters of $a,~b,~L$, while it is sensitive to the charge $Q$ as shown in Fig. \ref{lambda-Teff-Q}.

We assume that a static observe at radius coordinate $r_0$ sends light rays into the past, the angle $\alpha$ between such a light ray and the radial direction reads
\begin{eqnarray}
\cot^2\alpha=\frac{g_{rr}}{g_{\phi\phi}}\left(\frac{dr}{d\phi}\right)^2\bigg|_{r_0}
\end{eqnarray}
Considering eqs. (\ref{drphi}) and (\ref{bR}), the above equation becomes
\begin{eqnarray}
\sin^2\alpha=\frac{f(r_0)}{b^2r_0^2}. \label{alpha}
\end{eqnarray}
As we know that a black hole can capture all light falling onto in and it emits nothing. Therefore even a naive consideration suggests that an observer will see a dark spot in the sky where the black hole is supposed to be located. However, due to the strong bending of light rays by the black hole gravity, both the size and the shape of this spot are different from what we naively expect on the basis of Euclidean geometry from looking at a non-gravitating  black ball. In the case of a spherically symmetric black hole, the difference between the shadow and the imaginary Euclidean image of the black hole is only in angular size. The size and shape of the shadow depend not only on the parameters of the black hole itself, but also on the position of the observer \cite{Perlick2022}. The boundary curve of the shadow corresponds to pase-oriented light rays that asymptotically approach one of the unstable circular photon orbit at $r_{ph}$. Thus when $R\rightarrow r_{ph}$, we have the radius and the angular radius of the shadow as
\begin{eqnarray}
&&r_{sh}\equiv b_{cr}=r_{ph}/\sqrt{f(r_{ph})},\\
&&\sin^2\alpha_{sh}=\frac{r_{sh}^2f(r_0)}{r_0^2},\label{alphash}
\end{eqnarray}
which is consistent with the results in Ref. \cite{Perlick2022}. For an observer at a large distance, $f(r_0)\rightarrow\frac{1}{(1-a)}\left(1-\Lambda r_0^2/3\right)$, the angular radius of shadow in eq. (\ref{alphash}) can be approximated by
\begin{eqnarray}
\alpha_{sh}=\frac{r_{sh}\left(1-\Lambda r_0^2/3\right)}{(1-a)r_0}.
\end{eqnarray}

\begin{figure}[htp]
\includegraphics[width=0.55\textwidth]{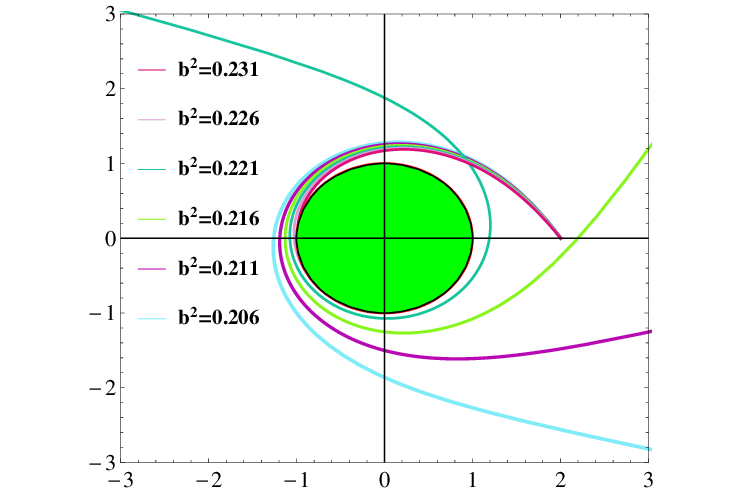}
\caption{The light rays with different values of the impact parameter $b$. The parameters are set to $a=0.0001$, $\Lambda=0.0857$, $M=0.5$, and $Q=0.5$.}\label{lr-KR}
\end{figure}

\begin{figure}[htp]
\subfigure[$~~r_{sh}-a$]{\includegraphics[width=0.4\textwidth]{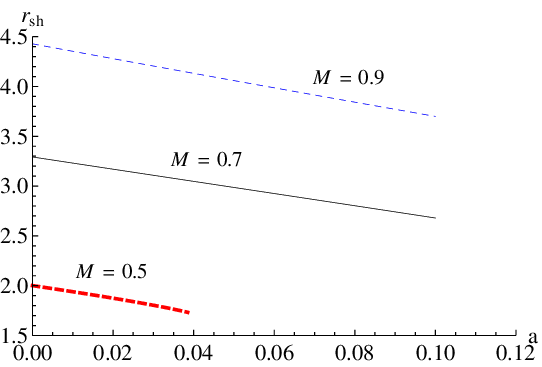}\label{rsha}}~~~~
\subfigure[$~~\alpha_{sh}-a$]{\includegraphics[width=0.4\textwidth]{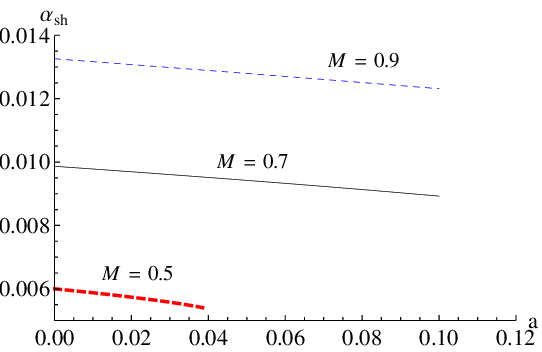}\label{alphaa}}~~~~
\caption{The horizon radius and the shadow radius vs the Lorentz-violating parameter. The parameters are set to $Q=0.5,~\Lambda=0.0006,~r_0=1000$.}\label{rsh-alpha-a}
\end{figure}

For the KR-dS black hole, we depict the light ray with different values of the impact parameter $b$ in Fig. \ref{lr-KR}, where the red circular represents the unstable photon orbit. The results show that the photons with the large impact parameter are trapped by the black hole, and only those photons with the small impact parameter can escape from it. The influence of the Lorentz-violating parameter on the radius and the angular radius of the shadow for the KR-dS black hole are illustrated in Fig. \ref{rsh-alpha-a}. The radius and the angular radius of the shadow for the KR-dS black hole are both decreases with the Lorentz-violating parameter, furthermore they are inversely proportional to the Lorentz-violating parameter.


\section{Conclusion}
\label{Five}

The static and spherically symmetric dS spacetime with a black hole was investigated in the gravity theory featuring a nonzero VEV of the KR filed. By introducing the interplay between the black hole horizon and the cosmological horizon, in the usual way we derived the thermodynamical effective parameters and investigated the thermodynamic phase transition that exhibit new properties resulting from the Lorentz-violating effect. Interestingly, the critical ratio between two horizons, horizon radius, volume, and entropy are only related to the charge parameter, instead of the Lorenz-violating parameter, while the critical effective temperature and pressure are determined by both two parameters $Q$ and $a$. Furthermore, the critical effective temperature and pressure are both increasing with the Lorentz-violating parameter $a$, while they are decreasing with the charge parameter $Q$ which is fully inverse to that in AdS black hole systems.

On the other hand we also probed the phase structure of this system by the way of the Lyapunov exponent. Especially, through the effective potential of the massless photon orbit we found the stability of this system can be characterized by the Lyapunov exponent. And in the diagram of $\lambda-T_{eff}$, there exists the non-monotonic behavior which means the existence of the first-order phase transition. This type phase transition is just the one in the diagram of $P_{eff}-V$ as shown in \cite{Lyu2023}. In order to explore the effect of Lorentz violation on the motion of photon near the black hole, we studied the light rays and the shadow of the massless photons for the KR-dS black hole. Our results demonstrated that the photons with the large impact parameter are trapped by the black hole, and only those photons with the small impact parameter can escape from it. And the shadow and angular radii of this type black hole decrease with the Lorentz-violating parameter $a$. In particular compared with the angular radius $\alpha$ of the shadow, the shadow radius $r_{sh}$ exhibits a high sensitivity to variations in $a$.

\section*{Acknowledgments}
This work was supported by the Natural Science Foundation of China a under Grant No. 12075143 and No. 12375050, the Natural Science Foundation of Shanxi Province, China under Grant No. 202203021221209 and No. 202303021211180, and the Teaching Reform Project of Shanxi Datong Universtiy with Grant No. XJG2022234.


\begin{references}

\bibitem{Bekenstein1974}J. D. Bekenstein, {\it Generalized second law of thermodynamics in black hole physics,} Phys. Rev. D 9 (1974), 3292-3300.
\bibitem{Bardeen1973}J. M. Bardeen, B. Carter, and S. W. Hawking, {\it The Four laws of black hole mechanics,} Commun. Math. Phys. 31 (1973), 161-170.
\bibitem{Hawking1983}S. W. Hawking and D. N. Page, {\it Thermodynamics of Black Holes in anti-De Sitter Space,} Commun. Math. Phys. 87 (1983), 577.
\bibitem{Cai2008}R.-G. Cai, Z.-Y. Nie, and Y.-W. Sun, {\it Shear Viscosity from Effective Couplings of Gravitons}, Phys. Rev. D 78 (2008) 126007, arXiv:0811.1665.
\bibitem{Wei2015}S.-W Wei and Y.-X Liu, {\it Insight into the microscopic structure of an AdS black hole from a thermodynammical phase transition}, Phys. Rev. Lett 115 (2015) 111302.
\bibitem{Zhang2015}J.-L Zhang, R.-G Cai, and H.-W Yu, {\it Phase transition and Thermodynamical geometry of Reissner-Nordstr\"{o}m-AdS Black Holes in Extended Phase Space}, Phys. Rev. D 91, 044028 (2015), arXiv:1502.01428.
\bibitem{Du2021}Y.-Z Du, H.-F Li, F. Liu, R. Zhao, and L.-C Zhang, {\it Phase transition of non-linear charged Anti-de Sitter black holes}, Chin. Phys. C 45 (2021) 11, arXiv:2112.10403.
\bibitem{He2010}X. He, B. Wang, R. G. Cai, and C. Y. Lin, {\it Signature of the black hole phase transition in quasinormal modes,} Phys. Lett. B 688 (2010), 230-236.
\bibitem{Yao2022}F. Yao and J. Tao, {\it Extended phase space thermodynamics for dyonic black holes with a power Maxwell field,} Phys. Rev. D 105 (2022) no.12, 124018.
\bibitem{Wei2022}S.-W Wei, Y.-X Liu, and R. B. Mann, {\it Black Hole Solutions as Topological Thermodynamics Defects,} Phys. Rev. L 129, 191101 (2022).
\bibitem{He2017}S. He, L.-F. Li, and X. X. Zeng, {\it Holographic Van der Waals-like phase transition in the Gauss¨CBonnet gravity,} Nucl. Phys. B 915 (2017), 243-261.
\bibitem{Huang2021}Y. Huang, H. Jing, J. Tao, and F. Yao, {\it Phase structures and transitions of quintessence surrounding RN black holes in a grand canonical ensemble,} Chin. Phys. C 45 (2021) 7, 075101.
\bibitem{Bai2022}N. Bai, A. He, and J. Tao, {\it Microstructure of charged AdS black hole with minimal length effects}, Chin. Phys. C 46 (2022) 12, 125105.
\bibitem{Wang2020}P. Wang, H. Yang, and S. Ying, {\it Thermodynamics and phase transition of a Gauss-Bonnet black hole in a cavity,} Phys. Rev. D 101 (2020) 6, 064045.
\bibitem{Hamil2023}B. Hamil and B. C. Lutfuoglu, {\it Thermodynamics and Shadows of quantum-corrected Reissner-Nordstr$\ddot{o}m$ black hole surrounded by quintessence,} Phys.Dark Univ. 42 (2023) 101293, arXiv:2305.07123.
\bibitem{Chamblin1999} A. Chamblin, R. Emparan, C. V. Johnson, and R. C. Myers, {\it Holography, Thermodydnamics and Fluctuations of Charged AdS Black Holes}, Phys. Rev. D 60 (1999) 104026, arXiv:hep-th/9904197.
\bibitem{Chamblin1999a}A. Chamblin, R. Emparan, C. V. Johnson, and R. C. Myers, {\it Charged AdS Black Holes and Catastrophic Holography}, Phys. Rev. D 60 (1999) 064018, arXiv:hep-th/9902170.
\bibitem{Dolan2013}B. P. Dolan, D. Kastor, D. Kubiznak, R. B. Mann, and J. Traschen, {\it Thermodynamic Volumes and Isoperimetric Inequalities for de Sitter Black Holes}, Phys. Rev. D 87 (2013) 104017, arXiv:1301.5926.
\bibitem{Tannukij2020}L. Tannukij, P. Wongjun, E. Hirunsirisawat, T. Deesuwan, and C. Promsiri, {\it Thermodynamics and phase transition of spherically symmetric black hole in de Sitter space from Renyi statistics,} Eur. Phys. J. Plus 135 (2020) 6, arXiv:2002.00377.
\bibitem{Chabab2021}M. Chabab, H. E. Moumni, and J. Khalloufi, {\it On Einstein-non linear-Maxwell-Yukawa de-Sitter black hole thermodynamics,} Nucl. Phys. B 963 (2021) 115305, arXiv:2001.01134.
\bibitem{Wei2019}Sh.-W Wei and Y.-X Liu, {\it Null Geodesics, Quasinormal Modes, and Thermodynamic Phase Transition for Charged Black Holes in Asymptotically Flat and dS Spacetimes,} Chin. Phys. C 44 (2020) 11, arXiv:1909.11911.
\bibitem{Simovic2019}F. Simovic and R. B. Mann, {\it Critical Phenomena of Charged de Sitter Black Holes in Cavities,} Class. Quant. Grav. 36 (2019) 1, arXiv:1807.11875.
\bibitem{Nam2018}C. H. Nam, {\it Non-linear charged dS black hole and its thermodynamics and phase transitions,} Eur.Phys.J.C 78 (2018) 5.
\bibitem{Kubiznak2016}D. Kubiznak and F. Simovic, {\it Thermodynamics of horizons: de Sitter black holes and reentrant phase transitions,} Class.Quant.Grav. 33 (2016) 24, arXiv:1507.08630.
\bibitem{Zhang2014}L.-Ch Zhang, M.-S Ma, H.-H Zhao, and R Zhao, {\it Thermodynamics of phase transition in higher-dimensional Reissner-Nordstr$\ddot{o}$m-de Sitter black hole,} Eur. Phys. J. C 74 (2014) 9, arXiv:1403.2151.
\bibitem{Cvetic2002}M. Cvetic, S. Nojiri, and S. D. Odintsov, {\it Black hole thermodynamics and negative entropy in de Sitter and anti-de Sitter Einstein-Gauss-Bonnet gravity,} Nucl. Phys. B 628 (2002) 295-330, arXiv:hep-th/0112045.
\bibitem{Zhao2021}H.-H Zhao, L.-Ch Zhang, and F. Liu, {\it Thermodynamics and phase transition of topological dS black holes with a nonlinear source,} Commun.Theor.Phys. 73 (2021) 9, arXiv:1704.05167.
\bibitem{He2018}Y. He, M.-S. Ma, and R. Zhao, {\it Entropy of black holes with multiple horizons}, Nucl. Phys. B 930, 513 (2018).
\bibitem{Ma2018}Y.-B Ma, L.-C Zhang, S. Cao, T.-H Liu, S.-B Geng, et al, {\it Entropy of the electrically charged hairy black holes}, Eur. Phys. J. C 78 (2018) 9, 763.
\bibitem{Zhao2018}R Zhao and L.-C Zhang, {\it Entropy of higher-dimensional charged de Sitter black holes and phase transition}, Commun. Theor. Phys. 70 (2018) 5, 578, arXiv:1710.07225.
\bibitem{Zhang2019}L.-C Zhang and R. Zhao, {\it The entropic force in Reissoner-Nordstr$\ddot{o}$m-de Sitter spacetime}, Phys. Lett. B 797 (2019) 10, 134798.
\bibitem{Liu2019}F. Liu and Li-Chun Zhang, {\it On thermodynamics of RN-dS black hole surrounded by the quintessence}, Chin. Journal of Phys 57, 2019, 53-60.
\bibitem{Guo2020}X.-Y Guo, H.-F Li, and L.-C Zhang, {\it Entropy of higher-dimensional charged Gauss¨CBonnet black hole in de Sitter space}, Commun. Theor. Phys. 72 (2020) 8, 085403, arXiv:1803.09456.
\bibitem{Gubser1998}S. S. Gubser, I. R. Klebanov, and A. M. Polyakov, {\it Gauge Theory Correlators from Non-Critical String Theory,} Phys. Lett. B 428, 105 (1998).
\bibitem{Maldacena1998}J. M. Maldacena, {\it The Large N Limit of Superconformal Field Theories and Supergravity}, Adv. Theor. Math. Phys. 2, 231 (1998).
\bibitem{Strominger2001}A. Strominger, {\it The dS/CFT correspondence}, JHEP 0110 (2001) 034.
\bibitem{Kostelecky1989}V. A. Kostelecky and S. Samuel, {\it Spontaneous breaking of Lorentz symmetry in string theory, }Phys. Rev. D 39 (1989) 683.
\bibitem{Alfaro2002}J. Alfaro, H. A. Morales-Tecotl, and L. F. Urrutia, {\it Loop quantum gravity and light propagation, }Phys. Rev. D 65 (2002) 103509, arXiv:hep-th/0108061.
\bibitem{Horava2009}P. Horava, {\it Quantum gravity at a Lifshitz point,} Phys. Rev. D 79 (2009) 084008, arXiv:0901.3775.
\bibitem{Kostelecky1989a}V. A. Kostelecky and S. Samuel, {\it Gravitational Phenomenology in Higher Dimensional Theories and Strings,} Phys. Rev. D 40 (1989) 1886.
\bibitem{Kostelecky1989b}V. A. Kostelecky and S. Samuel, {\it Phenomenological gravitational constraints on strings and higher dimensional theories,} Phys. Rev. Lett. 63 (1989) 224.
\bibitem{Bailey2006}Q. G. Bailey and V. A. Kostelecky, {\it Signals for Lorentz violation in post-Newtonian gravity,} Phys. Rev. D 74 (2006) 045001, arXiv:gr-qc/0603030.
\bibitem{Bluhm2008}R. Bluhm, N. L. Gagne, R. Potting, and A. Vrublevskis, {\it Constraints and stability in vector theories with spontaneous Lorentz violation,} Phys. Rev. D 77 (2008) 125007, arXiv:0802.4071.
\bibitem{Casana2018} R. Casana, A. Cavalcante, F.P. Poulis, and E.B. Santos, {\it Exact Schwarzschild-like solution in a bumblebee gravity model,} Phys. Rev. 97 (2018) 104001, arXiv:1711.02273.
\bibitem{Kanzi2019}S. Kanzi and I. Sakall, {\it GUP modified hawking radiation in bumblebee gravity,} Nucl. Phys. B 946 (2019) 114703, arXiv:1905.00477.
\bibitem{Maluf2021}R. V. Maluf and J. C. S. Neves, {\it Black holes with a cosmological constant in bumblebee gravity,} Phys. Rev. D 103 (2021) 044002, arXiv:2011.12841.
\bibitem{Cai2023}Z. Cai and R.-J. Yang, {\it Accretion of the Vlasov gas onto a Schwarzschild-like black hole,} Phys. Dark Univ. 42 (2023) 101292, arXiv:2205.04826.
\bibitem{Xu2023}R. Xu, D. Liang, and L. Shao, {\it Static spherical vacuum solutions in the bumblebee gravity model,} Phys. Rev. D 107 (2023) 024011, arXiv:2209.02209.
\bibitem{Ding2020}C. Ding, C. Liu, R. Casana, and A. Cavalcante, {\it Exact Kerr-like solution and its shadow in a gravity model with spontaneous Lorentz symmetry breaking,} Eur. Phys. J. C 80 (2020) 178, arXiv:1910.02674.
\bibitem{Ding2021}C. Ding and X. Chen, {\it Slowly rotating Einstein-bumblebee black hole solution and its greybody factor in a Lorentz violation model,} Chin. Phys. C 45 (2021) 025106, arXiv:2008.10474.
\bibitem{Wang2022}H.-M. Wang and S.-W. Wei, {\it Shadow cast by Kerr-like black hole in the presence of plasma in Einstein-bumblebee gravity,} Eur. Phys. J. Plus 137 (2022) 571, arXiv:2106.14602.
\bibitem{Liu2023}W. Liu, X. Fang, J. Jing, and J. Wang, {\it QNMs of slowly rotating Einstein-Bumblebee black hole,} Eur. Phys. J. C 83 (2023) 83, arXiv:2211.03156.
\bibitem{Altschul2010}B. Altschul, Q. G. Bailey, and V. A. Kostelecky, {\it Lorentz violation with an antisymmetric tensor,} Phys. Rev. D 81 (2010) 065028, arXiv:0912.4852.
\bibitem{Lessa2020}L. A. Lessa, J. E. G. Silva, R. V. Maluf, and C. A. S. Almeida, {\it Modified black hole solution with a background Kalb-Ramond field,} Eur. Phys. J. C 80 (2020) 335, arXiv:1911.10296.
\bibitem{Atamurotov2022}F. Atamurotov, D. Ortiqboev, A. Abdujabbarov, and G. Mustafa, {\it Particle dynamics and gravitational weak lensing around black hole in the Kalb-Ramond gravity,} Eur. Phys. J. C 82 (2022) 659.
\bibitem{Kumar2020}R. Kumar, S. G. Ghosh, and A. Wang, {\it Gravitational deflection of light and shadow cast by rotating Kalb-Ramond black holes,} Phys. Rev. D 101 (2020) 104001, arXiv:2001.00460.
\bibitem{Yang2023} K. Yang, Y.-Z Chen, Z.-Q Duan, and J.-Y Zhao, {\it Static and spherically symmetric black holes in gravity with a background Kalb-Ramond field}, Phys. Rev. D 108 (2023) 12, 124004, arXiv:2308.06613.





\bibitem{Suzuki1997}S. Suzuki and K. i. Maeda, {\it Chaos in Schwarzschild space-time: The motion of a spinning particle,} Phys. Rev. D 55 (1997), 4848-4859.
\bibitem{Lu2018}F. Lu, J. Tao, and P. Wang, {\it Minimal Length Effects on Chaotic Motion of Particles around Black Hole Horizon,} JCAP 12 (2018), 036.
\bibitem{Hartl2003}M. D. Hartl, {\it Dynamics of spinning test particles in Kerr space-time,} Phys. Rev. D 67 (2003), 024005.
\bibitem{Bombelli1992}L. Bombelli and E. Calzetta, {\it Chaos around a black hole,} Class. Quant. Grav. 9 (1992), 2573-2599.
\bibitem{Wang2017}M. Wang, S. Chen, and J. Jing, {\it Chaos in the motion of a test scalar particle coupling to the Einstein tensor in Schwarzschild-Melvin black hole spacetime,} Eur. Phys. J. C 77 (2017) 4.
\bibitem{Chen2016}S. Chen, M. Wang, and J. Jing, {\it Chaotic motion of particles in the accelerating and rotating black holes spacetime,} JHEP 09 (2016), 082.
\bibitem{Yang2023a}S. Yang, J. Tao, B. Mu, and A. He, {\it Lyapunov exponents and phase transitions of Born-Infeld AdS black holes}, JCAP 07 (2023), 045.
\bibitem{Synge1966}J. L. Synge, {\it The Escape of Photons from Gravitationally Intense Stars,} Mon. Not. Roy. Astron. Soc. 131 (1966) 3.
\bibitem{Vries1999}A. de Vries, {\it The apparent shape of a rotating charged black hole, closed photon orbits and the bifurcation set $A_4$,} Class. Quant. Grav. 17 (1999) 1.
\bibitem{Bardeen1972}J. M. Bardeen, W. H. Press, and S. A. Teukolsky, {\it Rotating black holes: Locally nonrotating frames, energy extraction, and scalar synchrotron radiation,} Astrophys. J. 178 (1972) 347.
\bibitem{Grenzebach2014}A. Grenzebach, V. Perlick, and C. Lammerzahl, {\it Photon Regions and Shadows of Kerr-Newman-NUT Black Holes with a Cosmological Constant,} Phys. Rev. D 89 (2014) 12.
\bibitem{Guo2018}M. Guo, N. A. Obers, and H. Yan, {\it Observational signatures of near-extremal Kerr-like black holes in a modified gravity theory at the Event Horizon Telescope,} Phys. Rev. D 98 (2018) 8.
\bibitem{Hennigar2018}R. A. Hennigar, M. B. J. Poshteh, and R. B. Mann, {\it Shadows, Signals, and Stability in Einsteinian Cubic Gravity,} Phys. Rev. D 97 (2018) 6.
\bibitem{Amir2018}M. Amir, B. P. Singh, and S. G. Ghosh, {\it Shadows of rotating five-dimensional charged EMCS black holes,} Eur. Phys. J. C 78 (2018) 5.
\bibitem{Jusufi2020}K. Jusufi, M. Jamil, and T. Zhu, {\it Shadows of Sgr $A^*$ black hole surrounded by superfluid dark matter halo,} Eur. Phys. J. C 80 (2020) 5.

\bibitem{Duan2023}Z.-Q Duan, J.-Y Zhao, and K. Yang, {\it Electrically charged black holes in gravity with a background Kalb-Ramond field}, arXiv:2310.13555.

\bibitem{Guo2022}X. Guo, Y. Lu, B. Mu, and P. Wang, {\it Probing phase structure of black holes with Lyapunov exponents}, JHEP 08 (2022), 153.

\bibitem{Cardoso2009}V. Cardoso, A. S. Miranda, E. Berti, H. Witek, and V. T. Zanchin, {\it Geodesic stability, Lyapunov exponents and quasinormal modes}, Phys. Rev. D 79 (2009) 6, 064016.

\bibitem{Wei2023}W. Wei and Y. X. Liu, {\it Aschenbach effect and circular orbits in static and spherically symmetric black hole backgrounds}, Phys.Dark Univ. 43 (2024) 101409, arXiv:2308.11883.



\bibitem{Lyu2023}X. Lyu, J. Tao, and P. Wang, {\it Probing the thermodynamics of charged Gauss Bonnet AdS black holes with the Lyapunov exponent}, arXiv:2312.11912.

\bibitem{Perlick2022}V. Perlick and O. Y. Tsupko, {\it Calculating black hole shadows: Review of analytical studies}, Phys. Rept. 947 (2022) 1-39, arXiv:2105.07101.








\end{references}
\end{document}